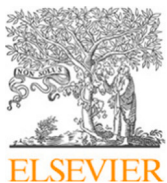
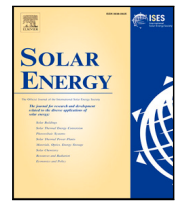

# Alternate stabilization methods for CZTSSe photovoltaic devices by thermal treatment, dark electric bias and illumination

W. Ananda [a,b,c,d,*], M. Rennhofer [a], A. Mittal [a,b,c], N. Zechner [a,e], W. Lang [b]

[a] *AIT Austrian Institute of Technology GmbH, Giefinggasse 2, 1210 Vienna, Austria*
[b] *Faculty of Physics, University of Vienna, Boltzmanngasse 5, 1090 Vienna, Austria*
[c] *Vienna Doctoral School in Physics, University of Vienna, Boltzmanngasse 5, 1090 Vienna, Austria*
[d] *Center for Material and Technical Product, Ministry of Industry, Sangkuriang 14, 40135 Bandung, Indonesia*
[e] *Institute of Energy Systems and Thermodynamics, Vienna University of Technology, Getreidemarkt 9, 1060 Vienna, Austria*

A R T I C L E   I N F O

A B S T R A C T

Reliable measurement routines are crucial for power rating and yield prediction of photovoltaic emerging thin-film technologies. Copper-Zinc-Tin-Sulfur-Selenium (CZTSSe) thin-film photovoltaic devices are an emerging technology made of abundant elements. Still, sufficient stabilization methods prior to electric power measurement are missing in the international standardization, while existing standards for other thin-film technologies do not work properly for CZTSSe. This study investigated methods for achieving power stabilization of the CZTSSe solar devices. Three complementary stabilization routines for the kesterite-based solar devices were investigated as an alternative to the existing international device testing standards: rapid annealing, dark electric biasing and different operating points under illumination. The typical number of stabilization cycles for power stabilization was between 3 and 6 cycles of rapid annealing, dark electric bias and illumination with a power loss of -19.5%, -11.4%, and -1.9%, for the respective methods. The dark electric bias method was found to provide the most reliable average result for power stabilization. All stabilization methods proved to have the potential to work sufficiently in stabilizing the CZTSSe devices for standardized power measurement.

## 1. Introduction

The development of solar photovoltaic (PV) technologies is becoming more desirable in recent years for dealing with the decarbonization pathways to fulfill the 1.5 °C scenario as mandated by Paris Agreement (COP 21, 2015; [IRENA] - International Renewable Energy Agency, 2022). Solar PV accounts for more than half of all renewable capacity additions in 2021, followed by wind and hydropower (International Energy Agency, 2021). The International Energy Agency (IEA) states that by 2050, solar PV could supply at least an estimated 16% of global electricity production (International Energy Agency, 2014). Several driving factors are the falling PV module price and scale flexibility of PV deployment (International Energy Agency, 2011).

Copper Zinc Tin Sulfur Selenium (CZTSSe) solar cells (also often referred to as kesterites) are seen by some as promising candidates for the next generation of thin-film (TF) PV because of their high theoretical efficiency, low-cost potential, and environmental friendliness (Redinger et al., 2011). These advantages are based on its significant absorption coefficient (>$10^4$ cm$^{-1}$, for a single-junction solar cell), well-aligned direct band gap of 1.45 eV, and its earth-abundance as well as being almost free of rare or toxic elements (Zhang et al., 2014). The elements of CZTSSe solar cells, Cu, Zn, Sn, S, and Se, are available in the earth's crust at approximately 50, 75, 2.2, 260, and 0.05 ppm (parts per million), respectively (Pal et al., 2019).

Despite the advantages of the CZTSSe technologies, there are still issues for improvement before market readiness. Besides the efficiency increase, which is a matter of material research, the reliability and assurance on a technical level have to be understood better. In this respect, one critical issue about CZTSSe solar devices for device application is determining its output power value and module stability (as it is for TF-PV in general), see e.g. Mittal et al. (2016), Rennhofer et al. (2017), Delahoy et al. (1998), Kenny et al. (2007) and Novalin et al. (2013). The module output power is not as stable under the standardized measurement as crystalline silicon PV modules (Rennhofer et al., 2017; Delahoy et al., 1998). The electric output power of thin-film photovoltaic devices has been, up till now, determined by the existing International Electrotechnical Commission (IEC) standards: (former IEC 61646) and IEC 61215-1-2–IEC 61215-1-4 (International Electrotechnical Commission, 2008, 2016a,b,d) after 2016, but there is still no explicit validated standard for the CZTSSe solar cells.

---

Reliable measurement of electrical parameters of TF-PV modules is crucial for power determination of devices under standard test conditions and total yield prediction of PV systems in the field. On the one side, an undefined state of order in a fresh PV device can lead to unstable behavior, increasing the error in Standard Test Conditions (STC) measurement. On the other hand, also metastable effects make it difficult to provide a reliable power rating of TF-PV modules (Kenny et al., 2007; Siebentritt et al., 2010). This can lead to inaccurate investment return calculations for installations of TF-PV systems and may also affect warranty cases (Menz, 2015). In copper-indium-gallium-selenide (CIGS) modules, metastable changes have been attributed to persistent photoconductivity due to copper migration in the bulk of CIGS (Guillemoles et al., 2000) and charging–discharging of defect states at the CdS/CIGS interface (Zabierowski et al., 2001). In cadmium-telluride (CdTe) modules, the motion of ion species, including copper, can affect metastable changes (Deline et al., 2012a).

The work presented here focused on measurement protocols and pre-conditioning procedures as fundamental prerequisites for power rating. The impact of measurement practice on standardized measurement routines (e.g. IEC) was analyzed. Basic work was also done in an international context prior to this study, where methods of alteration on the existing light-induced stabilization of Module Quality Tests (MQT) 19.1 in IEC 61215-2 on CIGS, CdTe, and CZTSSe technologies were investigated (Rennhofer, 2018).

Further understanding of the correlation between cell (i.e. device) properties and power measurement procedures will also allow to (a) evaluate cells of industrial production for their production quality and (b) quantify the damage from degradation or strongly degrading pre-conditioning protocols. Therefore, investigating alternative or adapted stabilization procedures is an important step to understand the reliable power and performance determination of modules.

The samples of this research were CZTSSe mini-modules. The samples were treated in different experimental routes where the environmental parameters of the pre-conditioning were altered. The pre-conditioning methods were low-temperature annealing, dark electric biasing, and controlled illumination.

## 2. Methodology

The module history, stabilization time, and measurement technique affect the thin-film modules' power stabilization. Existing IEC standard IEC 61215-1-x on technology-specific power stabilization of the thin-film modules is not always leading to optimal results as it is not adequate to balance or stabilize these effects. Therefore the main task of this research was to investigate the potential for stabilization of the CZTSSe solar devices by the above-mentioned means of low-temperature annealing, dark electric biasing, and controlled illumination. These represent the possible differing histories of modules fresh from the fabrication of the field. From the methodological approach, well-characterized samples of known initial states (all fresh from production) were put under well-controlled differing environmental conditions followed by Standard Test Conditions (STC) power rating according to the IEC standard. The STC for PV devices specifies a cell temperature of 25 °C and irradiance of 1000 Wm$^{-2}$ with an air mass of 1.5 (AM1.5) in a Static Solar Simulator Class AAA (International Electrotechnical Commission, 2016c).

### 2.1. Methodological approach

The chosen methods should be as less invasive as possible to be also applicable to fresh modules or in case of warranty. From the material scientific point of view, any stabilization routine may be an order–order (o–o), order–disorder (o–d), or disorder–order (d–o) transition.

From the point of view of optimal and stable behavior in operation, the methods supporting o–o or d–o transitions are preferable. From the point of view of repeatability and power labeling in the laboratory, all methods (also o–d transitions) may apply when allowing power determination with small power loss — and ideally being close or comparable to field operation values.

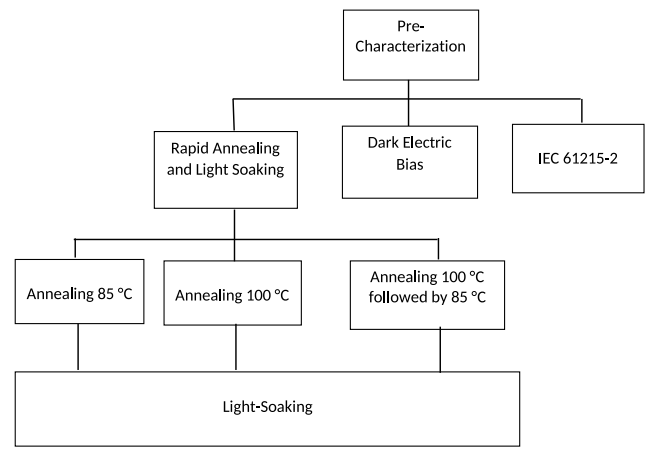

**Fig. 1.** The three main methods used in the experiments: rapid annealing, dark electric biasing, and illumination based on IEC 61215-2.

For the purpose of stabilization prior to power rating, physical parameters were chosen which may affect the order state in a different manner. These parameters are temperature, irradiance, and electrical current. The temperature is assumed to allow d–o or o–o transitions in low-temperature regime, i.e. affecting *isothermal annealing*. This can be understood as post-production ordering and should mainly increase the initial power. The irradiance is assumed to mainly drive o–o or o–d transitions as weak bonds (shallow states or van-der-Waals bonds) may be broken by photons, as also found for amorphous silicon (a-Si) (Lee et al., 2019). Finally, the effects of electrical forward bias (current) in the dark were reportedly different for CIGS and CdTe technologies. The use of forward current bias in the dark for CIGS modules can maintain the light-exposed state (Deline et al., 2012b). While in CdTe, the use of forward bias at elevated temperature without illumination did not consistently result in a light-exposed state (Silverman et al., 2015). The electric current may lead to d–o or o–o, but as well to o–d transitions. It is known that high currents (e.g. operation at short circuit current for a longer time) can damage thin films and interfaces in solar cells.

Following these parameters, three main methodologies were compared in this study as shown in Fig. 1. The first method is rapid annealing followed by a light soaking step. The second one is dark electric bias at three different temperatures. The third one is based on the IEC 61215-2 standard with some variations.

The formula used as stabilization criteria is based on standard IEC 61215-2 (International Electrotechnical Commission, 2016c):

$$\frac{(P_{max} - P_{min})}{P_{average}} < x \quad (1)$$

Here, $P_{max}$, $P_{min}$, and $P_{average}$ are defined as extreme values of three consecutive output power measurements taken from a sequence of alternating stabilization and measurement steps. The standard $x$ value for CZTSSe (as well as CIGS) solar devices is 2%.

After an initial power determination, the single stabilization methods are applied to the samples, and after each exposure step, the power is again determined by an STC measurement. Samples R1–R4 and D1–D6 were measured in a collimated light source using a class AAA static sun simulator for current–voltage ($I - V$) measurements at STC and light soaking. Samples L1–L3 were measured in a class AAA+ pulsed sun simulator for the $I - V$ measurement at STC, and in a static sun simulator for various standard and alternate light-induced stabilization procedures.

**Table 1**
Routes of the 3 main methods used in the experiments. R1–R4 are the routes of rapid annealing method with the total of 8 samples. Each route starts with annealing then followed by light soaking. D1–D6 are the routes of dark electric bias method with the total of 12 samples. Each route consists of 2 steps with the changing of $I_{SC}$ level and or the temperature. L1–L3 are the routes of illumination method with 2 samples of each route. This method mainly based on the standard of IEC 61215-2, with the 2 routes changing the temperature and operating point condition.

| Code | Routes | Samples |
| --- | --- | --- |
| R1 | Annealing 85 °C in dark for 48 h; followed by twice light-soaking each 4 h in 50 °C, 1000 Wm$^{-2}$, and under open-circuit condition. | R1-1, R1-2 |
| R2 | Annealing 85 °C in dark for 48 h; followed by twice light-soaking each 4 h in 25 °C, 1000 Wm$^{-2}$, and under open circuit condition. | R2-1, R2-2 |
| R3 | Annealing 100 °C in dark for 48 h; followed by twice light-soaking each 4 h in 25 °C, 1000 Wm$^{-2}$, and under open circuit condition. | R3-1, R3-2 |
| R4 | Annealing 100 °C and 85 °C in dark each for 24 h subsequently; then followed by twice light-soaking each 4 h in 25 °C, 1000 Wm$^{-2}$, and under open circuit condition. | R4-1, R4-2 |
| D1 | Dark bias in 1/3 $I_{SC}$ and 25 °C, followed by dark bias in 1/3 $I_{SC}$ and 50 °C. | D1-1, D1-2, D1-3 |
| D2 | Dark bias in 1/3 $I_{SC}$ and 25 °C, followed by dark bias in 2/3 $I_{SC}$ and 50 °C. | D2-1, D2-2 |
| D3 | Dark bias in 1/3 $I_{SC}$ and 50 °C, followed by dark bias in 2/3 $I_{SC}$ and 50 °C. | D3-1, D3-2 |
| D4 | Dark bias in 1/3 $I_{SC}$ and 50 °C, followed by dark bias in 1/3 $I_{SC}$ and 65 °C. | D4-1, D4-2 |
| D5 | Dark bias in 2/3 $I_{SC}$ and 50 °C, followed by dark bias in 2/3 $I_{SC}$ and 65 °C. | D5-1 |
| D6 | Dark bias in 2/3 $I_{SC}$ and 50 °C, followed by dark bias in 1 $I_{SC}$ and 50 °C. | D6-1, D6-2 |
| L1 | Standard pre-conditioning at MPP in 50 °C based on test requirement of IEC 61 215-2 | L1-1, L1-2 |
| L2 | Pre-conditioning based on test requirement of IEC 61 215-2 in 50 °C under open circuit condition | L2-1, L2-2 |
| L3 | Pre-conditioning based on test requirement of IEC 61 215-2 at MPP in 25 °C | L3-1, L3-2 |

### 2.1.1. Rapid annealing

This method was performed for triggering structural o–o, o–d or d–o-phenomena within the fresh-made samples to speed up the stabilization usually done only by medium temperatures (45 °C–55 °C) and illumination. The goal was to minimize the exposure steps needed for standard stabilization of IEC standards. The experiments used rapid annealing at elevated temperatures in the dark, followed by light soaking. Code R1–R4 in Table 1 shows the four different routes of this method using two annealing temperatures: 85 °C and 100 °C, respectively. The illumination or light-soaking was done twice for each route with the sample kept in dark storage in between. The STC $I-V$ measurement was done after annealing and after each light soaking.

### 2.1.2. Dark electric bias

Current induced changes are one candidate for relaxing a sample to stabilization, so dark bias was chosen for sample treatment. Forward bias in the dark has been proposed as a method to preserve the light-exposed electrical state (Dittmann et al., 2014; Del Cueto et al., 2010). As the kind of transition (mainly o–o or stabilizing and o–d or degrading) is assumed to be strongly current level dependent, the current levels were chosen between short-circuit currents ($I_{SC}$) of 0.3 $I_{SC}$ and 1.0 $I_{SC}$ of the individual specimen. During biasing the samples were kept in the darkness to suppress the effects of illumination. The relatively low starting bias current of 0.3 $I_{SC}$ was chosen to avoid interaction of the process of stabilization changes (o–o) with the systemic current-induced degradation of samples (o–d). The current bias was applied for 30 min. After 30 min, the bias current was turned off and the temperature was brought back to 25 °C for an STC $I-V$ measurement, completing one cycle. Measurements were carried on consecutively until the stabilization criterion (Eq. (1)) was met for three consecutive cycles.

The stabilization cycles were done at temperatures between 25 °C and 65 °C, respectively. Each of the modules underwent two rounds of biasing, rounds 1 and 2, each consisting of a full stabilization – at least three cycles – at this temperature. For the second round, there was a variation in temperature and/or current bias. The time between two rounds of biasing was always at least 24 h. This form of biasing in two steps was performed to separate the impact of higher temperatures from the impact of higher bias current. Code D1–D6 in Table 1 shows the pairs of temperature and current for each of the dark bias routes.

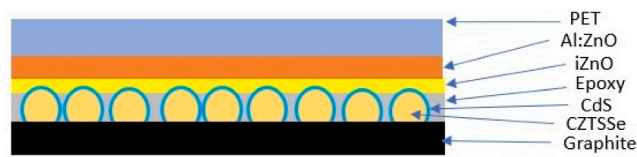

**Fig. 2.** The schematic diagram represent the CZTSSe mono-grain solar cell, where the mono-grains (CZTSSe grains coated with CdS buffer layer) are embedded in an epoxy layer. The graphite acts as a back contact and Al:ZnO layer acts as a front contact.

### 2.1.3. Illumination

For illumination stabilization, the experiments were carried out based on both standardized and non-standardized illumination routes to investigate optimized operation schemes. In the first case, the samples were stabilized according to IEC 61215-1-4 and IEC 61215-2 at their maximum-power-point (MPP) and 1000 Wm$^{-2}$. In the second and third cases, the samples were stabilized under light but changing the operating point ($V_{OC}$) or the temperature (25 °C). Each time the other parameters were kept according to IEC 61215-2. The three routes of this method can be seen as code L1–L3 in Table 1.

### 2.2. Material system and samples

The CZTSSe samples used for the experiments use mono-grain powder technology. In the growth of kesterite-type mono-grain powders, the crystals are formed in the presence of the liquid phase of water-soluble flux salts (KI, NaI, CdI$_2$) in an isothermal recrystallization process (Mellikov et al., 2007). A fine-tuning of the temperature, duration, and the specific flux material increases the homogeneity of the product absorber material and able to control the particle shape and size (Mellikov et al., 2015). The schematic diagram of the CZTSSe mono-grain solar cell is shown in Fig. 2.

Kesterite materials have a direct bandgap with an absorption coefficient of >10$^4$ cm$^{-1}$, suitable for thin-film PV applications. The band-gap is between 1.1 eV to 1.5 eV. Due to its structure, many lattice defects occur like vacancies, self-interstitials, anti-sites, and defect complexes. Due to these points defects, kesterites are showing high carrier recombination (Rey et al., 2016).

Theoretical calculations have revealed that the kesterite structure is prone to having cation disordering-related defects such as CuZn, CuSn, ZnCu, ZnSn, SnCu, and SnZn, and related thermodynamically stable defect clusters (He et al., 2021). Experimental studies identify the formation of intrinsic point defects near the front interface and within bulk (bulk recombination due to deep defects) as the key culprits behind the undesirable performance of solar cells (He et al., 2021). Teeter et al. conducted research about controlling metastable native point-defect populations in CIGS and CZTSSe materials and solar cells through voltage-bias annealing (Teeter et al., 2017).

Table 2
Initial $I-V$ parameters of all samples. The three samples: R1-1, D4-1, and L1-2 are chosen as the representative samples for each main method.

| No | Sample ID | $V_{OC}$ (V) | $I_{SC}$ (A) | $FF$ (%) | $P_{MPP}$ (W) |
|---|---|---|---|---|---|
| 1 | **R1-1** | **3.37** | **0.04** | **57.93** | **0.07** |
| 2 | R1-2 | 3.35 | 0.04 | 57.29 | 0.07 |
| 3 | R2-1 | 3.36 | 0.04 | 57.56 | 0.07 |
| 4 | R2-2 | 3.36 | 0.03 | 57.54 | 0.07 |
| 5 | R3-1 | 3.36 | 0.04 | 56.34 | 0.07 |
| 6 | R3-2 | 4.69 | 0.04 | 52.37 | 0.09 |
| 7 | R4-1 | 4.67 | 0.03 | 38.87 | 0.06 |
| 8 | R4-2 | 3.37 | 0.04 | 56.97 | 0.07 |
| 9 | D1-1 | 9.94 | 0.07 | 39.61 | 0.27 |
| 10 | D1-2 | 10.13 | 0.07 | 41.65 | 0.3 |
| 11 | D1-3 | 4.89 | 0.04 | 53.07 | 0.1 |
| 12 | D2-1 | 10.12 | 0.07 | 44.5 | 0.33 |
| 13 | D2-2 | 4.86 | 0.03 | 39.6 | 0.06 |
| 14 | D3-1 | 10.11 | 0.07 | 33.57 | 0.23 |
| 15 | D3-2 | 4.18 | 0.04 | 53.71 | 0.08 |
| 16 | **D4-1** | **4.88** | **0.04** | **57.18** | **0.1** |
| 17 | D4-2 | 4.92 | 0.03 | 55.85 | 0.09 |
| 18 | D5-1 | 10.38 | 0.07 | 56.83 | 0.42 |
| 19 | D6-1 | 4.87 | 0.04 | 58.09 | 0.1 |
| 20 | D6-2 | 10.41 | 0.07 | 54.77 | 0.43 |
| 21 | L1-1 | 10.35 | 0.07 | 56.47 | 0.44 |
| 22 | **L1-2** | **10.35** | **0.07** | **57.71** | **0.44** |
| 23 | L2-1 | 10.09 | 0.08 | 54.81 | 0.43 |
| 24 | L2-2 | 10.08 | 0.08 | 54.91 | 0.44 |
| 25 | L3-1 | 10.42 | 0.07 | 57.29 | 0.43 |
| 26 | L3-2 | 10.39 | 0.08 | 56.69 | 0.45 |

Table 3
Relative changes (%) of the initial I-V parameters for all samples. The three samples: R1-1, D4-1, and L1-2 are chosen as the representative samples for each main method.

| No | Sample ID | $V_{OC}$ | $I_{SC}$ | $FF$ | $P_{MPP}$ |
|---|---|---|---|---|---|
| 1 | **R1-1** | **−1.50** | **−1.11** | **−15.14** | **−17.34** |
| 2 | R1–2 | −0.81 | 0.32 | −10.44 | −10.87 |
| 3 | R2–1 | −0.68 | −0.85 | −11.40 | −12.76 |
| 4 | R2–2 | −0.40 | −1.47 | −13.87 | −15.47 |
| 5 | R3–1 | −0.98 | −1.27 | −23.34 | −25.05 |
| 6 | R3–2 | −4.56 | −2.97 | −18.69 | −24.71 |
| 7 | R4–1 | −2.99 | −5.14 | 2.65 | −5.54 |
| 8 | R4–2 | −0.82 | −1.99 | −28.21 | −30.22 |
| 9 | D1–1 | −1.16 | −3.46 | −2.04 | −6.52 |
| 10 | D1–2 | −2.86 | −5.43 | −5.35 | −13.06 |
| 11 | D1–3 | −0.96 | −2.84 | −5.05 | −8.61 |
| 12 | D2–1 | −6.01 | −1.14 | −6.36 | −13.00 |
| 13 | D2–2 | −5.17 | 1.59 | −2.75 | −6.24 |
| 14 | D3–1 | −2.95 | −8.20 | −1.31 | −12.05 |
| 15 | D3–2 | −0.14 | 0.49 | −1.32 | −0.97 |
| 16 | **D4-1** | **−1.28** | **−0.46** | **−5.53** | **−7.14** |
| 17 | D4–2 | −1.20 | −0.32 | −5.07 | −6.52 |
| 18 | D5–1 | −3.87 | −0.19 | −9.61 | −13.28 |
| 19 | D6–1 | −0.60 | 2.08 | −3.20 | −1.75 |
| 20 | D6–2 | −0.71 | −2.27 | −2.56 | −5.47 |
| 21 | L1–1 | −1.73 | 5.56 | −5.34 | −1.81 |
| 22 | **L1-2** | **−1.30** | **6.52** | **−2.63** | **2.38** |
| 23 | L2–1 | −1.70 | 1.77 | −8.11 | −8.07 |
| 24 | L2–2 | −1.44 | 0.80 | −7.18 | −7.78 |
| 27 | L3–1 | −1.35 | 8.95 | −2.45 | 4.84 |
| 28 | L3–2 | −2.28 | 5.55 | −3.91 | −0.89 |

The surface defect layer (SDL) formed between the cadmium-sulfide (CdS) and the CZTSSe accounts for interfacial defects. The band alignment mismatch can result in a barrier to the generated charge carriers or the recombination through traps. At the back contact the interaction of the absorber material with Mo results in the layer of $MoS_2$ or $MoSe_2$, which hinders the flow of the charge carriers through the back contact (Cernivec et al., 2011).

All these defects severely affect the $V_{OC}$ and the overall performance changes of the photovoltaic devices as well as stability changes under changed external conditions, see e.g. Mitzi et al. (2013). While the defects play a major role in the changes of electric parameters over time and thus are to an extent controlling the outcome of STC-power ratings, the weight of this work is on the stabilization routines followed by STC-power rating. Therefore, no direct defect measurements were done, while the understanding of the presence and correlation is supporting the interpretation of results while stabilizing samples.

Table 1 shows the total of 28 samples observed in this research, split into 3 methods: 8 samples for the rapid annealing, 12 samples for the dark current bias, and 8 samples for the IEC 61215 related methods. There are 3 types of CZTSSe mini-modules used in the experiments which came from different fabrication methods making the results more general. The first type consists of 2 dimensions: $5 \times 5\,cm^2$ and $9.5 \times 9\,cm^2$. The second type is $9.5 \times 9\,cm^2$. The third type size is $9.8 \times 8.9\,cm^2$. The efficiency of samples range from 4.0% until 5.1%. The initial sample parameters of all samples are given in Table 2.

## 3. Results and discussion

For each method, the results for all samples were accumulated and averaged. Then the sample that best shows the average behavior was selected for exemplary presentation of the time-dependent behavior in this section. Remarkable deviating behavior is mentioned still. The detailed results for all samples are summed up in Table 3 and Fig. 3. They are described and discussed in detail in the following sections.

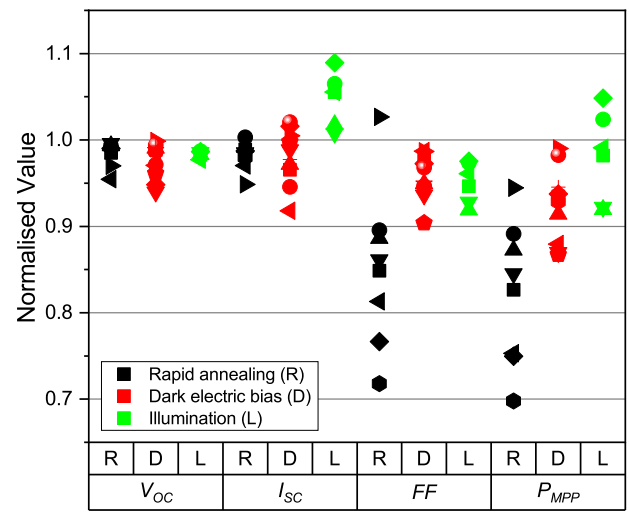

**Fig. 3.** Final value of four $I-V$ parameters of all samples. This graph shows that the homogeneity of $I-V$ parameters from high to low are: illumination method, dark current bias, and rapid annealing. $V_{OC}$ is the least affected parameter amongst all. Symbol with the same shape and color represents the same sample ID. (For interpretation of the references to color in this figure legend, the reader is referred to the web version of this article.)

### 3.1. Results for rapid annealing

All samples from the rapid annealing method, with the exception of sample R4-1, show comparable behavior (see Table 3 and Fig. 3). Power at MPP ($P_{MPP}$) shows the highest degradation followed by fill factor ($FF$) with the average value of −17.74% and −14.8% respectively. The $I_{SC}$ shows a higher degradation than the $V_{OC}$, especially after the first annealing step. The $V_{OC}$ was the most stable parameter among the four parameters observed. It was also found a recovery (increase) of the $V_{OC}$ and $I_{SC}$ values after degradation in the first annealing step.

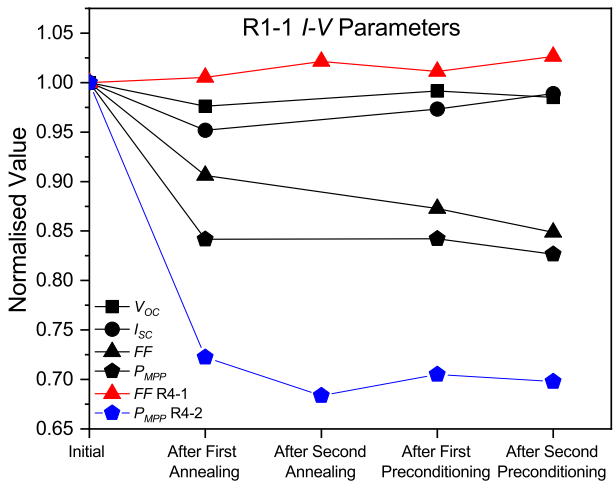

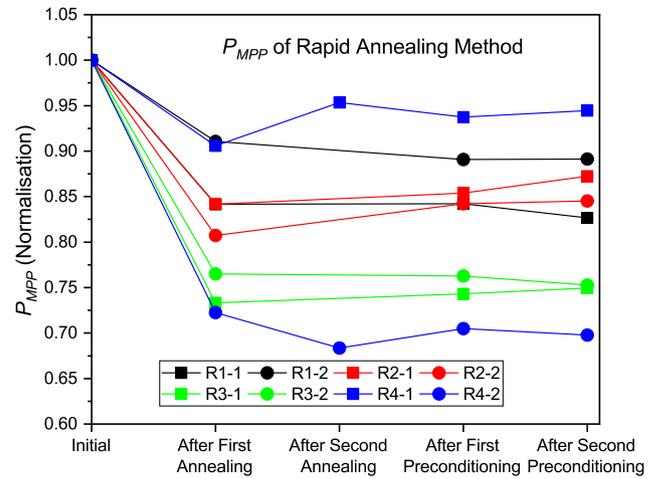

**Fig. 4.** The solid black symbols shows the changes of $V_{OC}$, $I_{SC}$, $FF$, and $P_{MPP}$ of the representative sample R1-1 for all rapid annealing steps. Also, the non-degrading $FF$ from sample R4-1 and the most degrading $P_{MPP}$ of R4-2 are shown. The lines are for guidance of the eyes.

**Fig. 5.** Change of $P_{MPP}$ of all samples during the route of the rapid annealing method. All samples show power degradation after the first annealing, followed by an almost constant $P_{MPP}$ afterwards. The lines are for guidance of the eyes.

The samples which underwent the annealing of 100 °C tend to have higher degradation of $I_{SC}$, $FF$, and $P_{MPP}$ than the one at 85 °C annealing. Sample R3-1, R3-2, and R4-2 which were annealed in 100 °C degraded in $P_{MPP}$ more than 20%, in $FF$ more than 15%, and in $I_{SC}$ more than 5% after the first annealing step. All samples from the R1 and R2 routes, which were annealed in 85 °C show smaller degradation values in these 3 parameters.

The results for all electric parameters of the representative sample R1-1 are shown in Fig. 4 together with remarkable parameters changes of other samples. For one sample also $FF$ was recovering while the rest of all $FF$ and also the $FF$ of the representative sample R1-1 are degrading. In general, the parameters of R1-1 show behavior as the mean parameter values do.

Fig. 5 shows the change of $P_{MPP}$ during the applied routes of the rapid annealing method. Samples from the same method show a relatively similar degradation and stabilization behavior between the two of them, except for route R4. The mean number of stabilization cycles was 3.25. After the second annealing, both samples from route R4 show different directions in a considerably opposite slope. This reveals that multiple annealing temperatures in one sample do not improve the reproducibility of the stabilization procedure.

Finally the power variations are tested versus formula (1), as shown in Fig. 6. Except for route 4, all other routes show a comparable result for the two samples in each route, respectively, with a difference in $P_{MPP}$ degradation values below 1%. Route R3 produces the most stabilized samples with the average $P_{MPP}$ value of 1.9%, followed closely by route R1 with 2.02%.

### 3.2. Results for dark electric biasing

Similar to the rapid annealing method, here $P_{MPP}$ is also the most degraded parameter (see Table 3 and Fig. 3). A number of 4 from 12 samples show $P_{MPP}$ degradation of more than 10%. The other parameters show less than 10% degradation for all samples. Since this method used current biasing, it is of interest whether $I_{SC}$ degrades, too. The result shows that $I_{SC}$ is the least affected parameter with an average degradation value of −1.68%. Nevertheless, there are 5 out of 12 samples, D1-1, D1-2, D1-3, D3-1, and D6-2, which show a higher relative loss of $I_{SC}$ compared to $V_{OC}$. The results for sample D4-1 are representing the general behavior of all samples, as can be seen in Fig. 7.

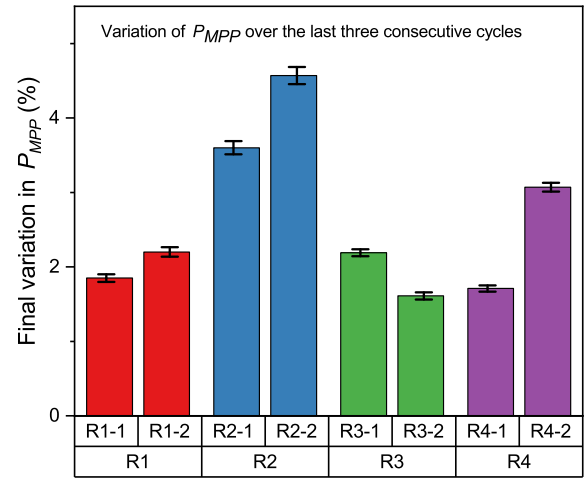

**Fig. 6.** Variation of $P_{MPP}$ over the last three consecutive cycles for all samples of the rapid annealing method based on Eq. (1). The standard deviation error bar is also shown for each sample.

Fig. 8 shows that route D4 has the best repeatability of the $P_{MPP}$ measurements because it shows similar behavior on both samples during the treatment. When compared to other routes, samples in route D4 have the closest value between them from the initial until the final stage of treatment. The mean number of stabilization cycles was 4.25.

Finally, the power variations of the dark electric bias method are tested versus formula (1), as shown in Fig. 9. This figure shows that among 12 samples, only 3 samples are above 2%: D1-1, D1-3, and D6-2. Route D4 turns out as the best stabilizing one with both specimens exhibiting the lowest average power variations of 0.96%. The result from both Figs. 8 and 9 suggest that a temperature higher than 25 °C and also a current bias <0.66 $I_{SC}$ is leading to a better stabilization procedure.

### 3.3. Results for illumination stabilization

Unlike the two previous methods, the parameter with the biggest average degradation compared to the initial values was found to be the $FF$ with −4.94% (see Table 3 and Fig. 3). Also, unlike the previous methods, all samples exhibited performance increase in some parameters, especially the $I_{SC}$. The $I_{SC}$ increased as high as 8.95% for sample L3-1. The illumination method showed the best performance in general with no parameter degradation more than 10% in all samples investigated. The results for sample L1-2 are representing the general behavior of all samples, as can be seen in Fig. 10.

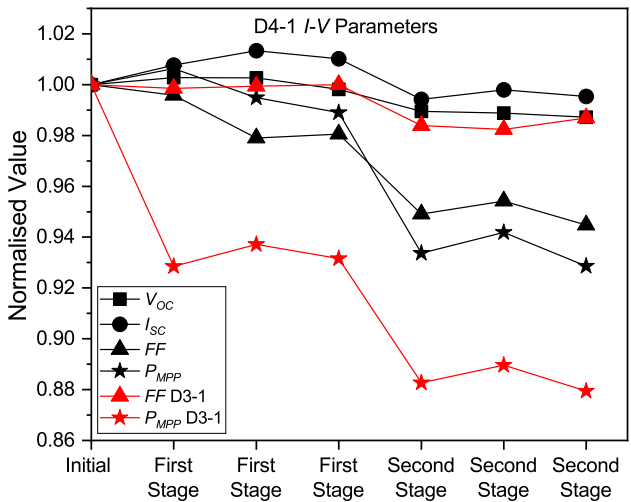

**Fig. 7.** The solid black symbols shows the changes of $V_{OC}$, $I_{SC}$, $FF$, and $P_{MPP}$ of the representative sample D4-1 for all dark electric bias steps. It also shows the anomaly of sample D3-1 which has the best $FF$ performance in this method albeit having the second worst $P_{MPP}$ performance. The lines are for guidance of the eyes.

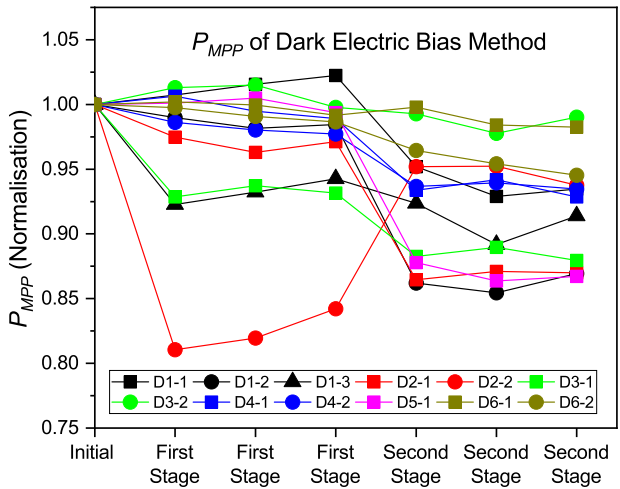

**Fig. 8.** Changing of $P_{MPP}$ for all samples during the route of dark bias method. The lines are for guidance of the eyes.

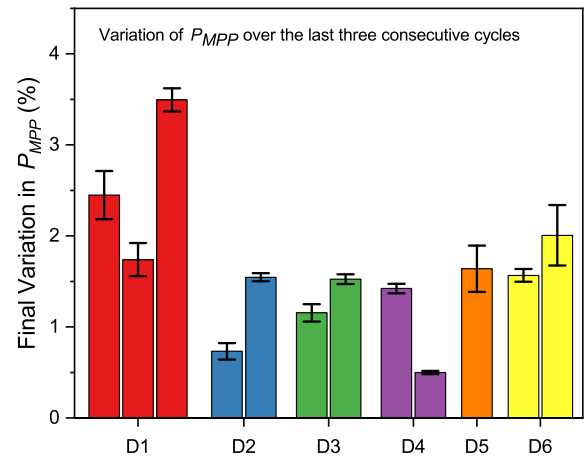

**Fig. 9.** Variation of $P_{MPP}$ over the last three consecutive cycles for all samples of the dark electric bias method based on Eq. (1). The standard deviation error bar is also shown for each sample.

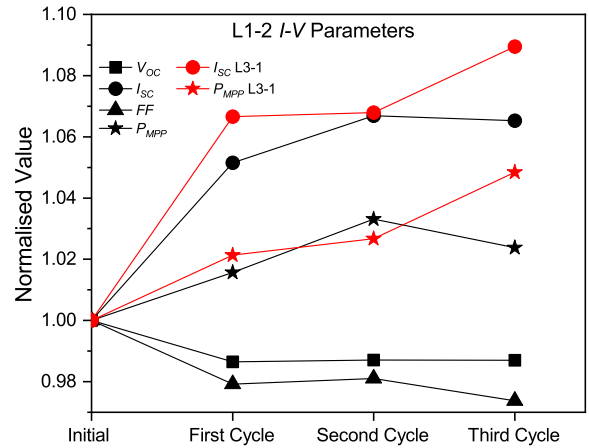

**Fig. 10.** The solid black symbols show the changes of $V_{OC}$, $I_{SC}$, $FF$, and $P_{MPP}$ of the representative sample L1-2 for all illumination method steps. It also shows the highest increase of $I_{SC}$ until 8.95% and $P_{MPP}$ until 4.84% in sample L3-1. The lines are for guidance of the eyes.

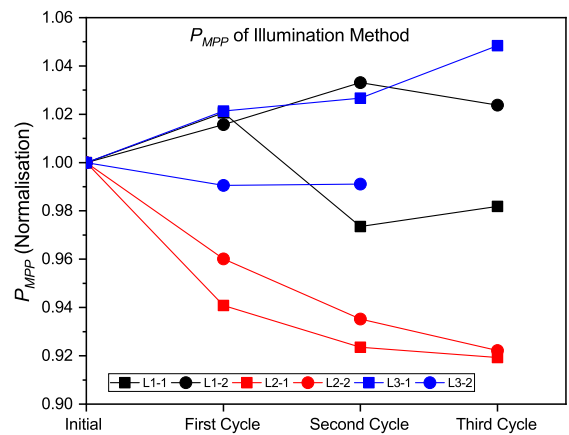

**Fig. 11.** Change of $P_{MPP}$ for all samples during the route of illumination method. The lines are for guidance of the eyes.

Fig. 11 shows that both samples from route L1 show different behavior during the treatment, especially after the first cycle. It also shows that both samples from route L2 have the highest $P_{MPP}$ degradation with values >7%. The two samples from route L3 show appealing results. One sample shows stable $P_{MPP}$ values, while the other one shows a continuous increase of $P_{MPP}$ with repeated illuminations. The mean number of stabilization cycles was 3.17.

Fig. 12 shows that route L1 which is the standard stabilization procedure based on IEC 61215-2 does not yield the best result for the CZTSSe samples tested. Further, the results for $P_{MPP}$ of the two specimens in route L1 were not comparable. It was found that the second sample in route L1 exhibits a much better result than the first sample. Route L1 also shows the largest variation of the $P_{MPP}$ values with a standard deviation of 3.22%. Route L3 on the other hand, operating at 25 °C during illumination, shows the best stabilization with an average $P_{MPP}$ variation of 1.79%.

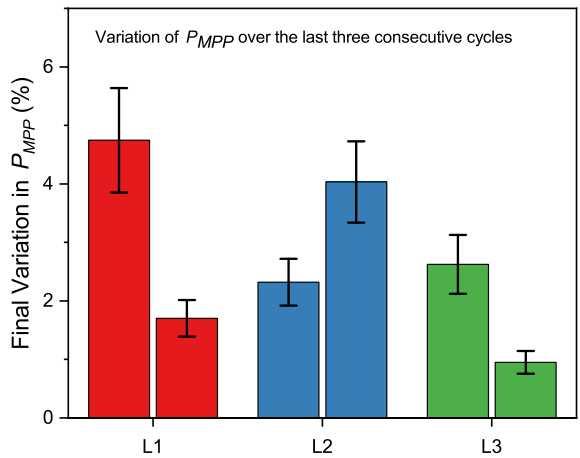

**Fig. 12.** Variation of $P_{MPP}$ over the last three consecutive cycles for all samples of the illumination method based on Eq. (1). Error bars represent the standard deviation for each sample.

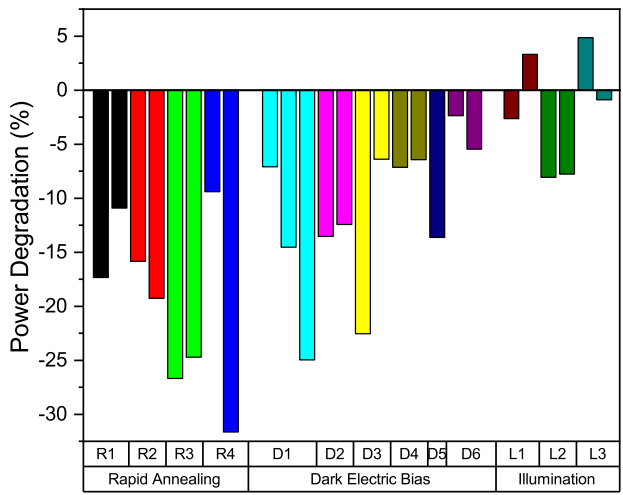

**Fig. 13.** Comparison of power degradation from 3 main stabilization procedures. This figure shows the large difference of the initial $P_{MPP}$ as compared to the power delivery during the experiments.

### 3.4. Discussion

The summarized results for power degradation of all samples are given in Fig. 13. The summarized behavior corresponding to the IEC-stabilization criterion are given in Fig. 14 for the representative samples and the best samples of each method, respectively.

#### 3.4.1. Rapid annealing

Depending on activation energy, defect structure, sample history (e.g. film deposition temperature and cooling rate), and annealing temperature the method may lead to degrading or ordering effects. These may be driven e.g. by solid-state isothermal annealing or post ordering by annihilating of point defects (Siebentritt et al., 2010; Rey et al., 2016). The device improvement by annealing has been reported by several research papers (He et al., 2021; Teeter et al., 2017). The improvement of device performance was mainly reported for annealing temperature >100 °C (Teeter et al., 2017; Su et al., 2020).

The annealing at $T \leq 100$ °C in our studied devices in general decreased the device performance. Both $P_{MPP}$ and $FF$ degrade the strongest with average >10%, followed by −1.81% average degradation of $I_{SC}$. These findings are compatible with the CZTSSe defect structure, as discussed in Section 2.2. This allows for dynamics arising both from bulk defects and heterojunction interfaces, e.g., to the transparent conducting oxide (TCO) window layer (Terlemezoglu et al., 2019). The aluminum-doped zinc oxide (AZO) window layers (like for the devices of this study) lead primarily to disordering dynamics, compared to other types in use, see Su et al. (2020) and Theelen and Daume (2016).

From the illumination step following each rapid annealing a contribution of illumination-driven mechanisms has to be considered also. For the CZTSSe specimens in this investigation, the implementation of 100 °C annealing or light soaking at 50 °C according to the IEC 61215-2 standard performed better than the combination of 85 °C annealing and 25 °C light soaking.

#### 3.4.2. Dark electric bias

The forward-bias current injection may have a stabilizing effect assuming that the current-injected charge carriers act as an equivalent to photo-carriers created by continuous light exposure (Deline et al., 2012b). Based on Table 3, apart of the $P_{MPP}$, the primary degradation of $I-V$ characteristics was a reduction of the $FF$ (−4.18% in average), followed by $V_{OC}$ (−2.24% in average). Deline et al. (2012b) also found that the primary $I-V$ parameter showing losses during dark exposure was the $FF$ followed by $V_{OC}$, which is conclusive with the proposed defect mechanism in CIGS by Deline et al. (2012b) and Lany and Zunger (2006). In other studies, very diverse results were found for dark bias application for stabilization, ranging from device improvement (Deline et al., 2012a) to various different results for stabilization in CIGS devices (Novalin et al., 2013).

Based on the assumption that forward-bias current injection is not fully equivalent to illumination-based charge carrier generation and that also specimens of the same production routes show high fluctuation in results and systematic behavior, the method has to be handled with care (Deline et al., 2012b; Silverman et al., 2015). This also minimizes its possibly high potential for automated and fast stabilization in an alternate IEC standard.

#### 3.4.3. Illumination

Table 3 shows the degradation of $FF$ and $V_{OC}$ in all samples with an average of −4.94% and −1.63% respectively for this method including the one according to the IEC standard. The $I_{SC}$ on the other hand, shows performance improvement with an average of 4.86%. Structural changes originating from illumination might be caused by effects (e.g. defects in the buffer layer) similar to the CIGS technology (Jones, 2020).

One contribution to the $V_{OC}$ reduction could be based in the existence of band tailing due to meta-stable electric behavior (Gokmen et al., 2013). Nikolaeva et al. managed to decrease the amplitudes of the investigated effect in CIGS solar cells by doing one-cycle 30 min of illumination and one week in the dark, leading to an increase of $V_{OC}$ (Nikolaeva et al., 2020). The cycles in our study took much longer than 30 min and at least 3 consecutive cycles were conducted (except sample L3-2). The second route of the illumination method, which is conducted under open-circuit condition, also showed the $V_{OC}$ rise after the first cycle but degraded later. The other two routes exhibited $V_{OC}$ degradation since the first cycle.

#### 3.4.4. Comparison of methods

Based on Table 3 and Fig. 3, the rapid annealing method causes the highest degradation in three parameters: $I_{SC}$, $FF$, and $P_{MPP}$; but this method has a slightly better $V_{OC}$ performance amongst all. The dark electric bias method has the highest degradation of $V_{OC}$ and the lowest $FF$ degradation. Finally, the illumination method has the best average performance of $I_{SC}$ and $P_{MPP}$.

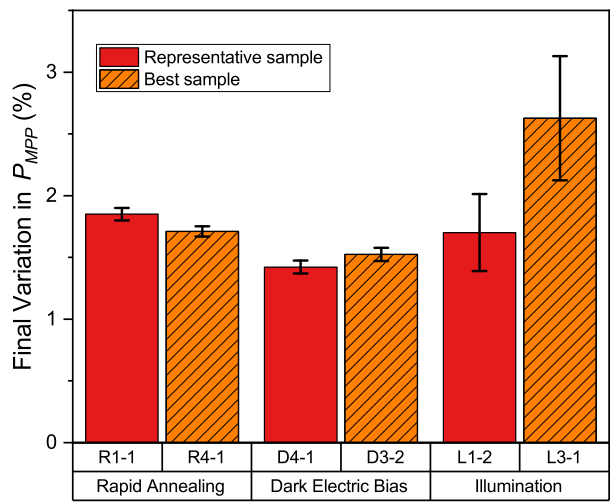

**Fig. 14.** Variation of $P_{MPP}$ over the last three consecutive cycles for the representative and best sample of each method based on Eq. (1). The best sample is chosen based on the lowest average degradation value across four $I–V$ parameters ($V_{OC}$, $I_{SC}$, $FF$, and $P_{MPP}$), not based on the stabilization criteria as on Eq. (1). That is why it is possible the least degraded sample is not the most stabilized.

Fig. 3 shows that, except for one sample, there are no improvements both in $V_{OC}$ and $FF$ by methods involving white light-soaking, e.g. rapid annealing and illumination. This is in contrast to CIGS solar cells' behavior which improves the $V_{OC}$ and $FF$ after white light-soaking (Siebentritt et al., 2010). The Cu-Zn disorder, particularly in CZTSSe, could be the cause of voltage losses which in theory is inherently difficult to reduce due to the slow kinetics of ordering (Scragg et al., 2016).

Fig. 13 shows the comparison of the power changes of all samples during the stabilization routes applied. Route R1 gives the lowest average power degradation in the rapid annealing method. This route is the only route that used 50 °C light soaking in the rapid annealing method. Six samples treated with the dark electric bias method degraded as little as using the illumination method. All samples from routes D4 and D6 in particular, are among those 6 samples. In the illumination method, route L3 shows the lowest power degradation followed by route L1. In general, the illumination method gives the lowest power degradation of all.

In Fig. 14 fulfillment of the power stabilization criteria according to IEC is compared for the representative samples and the best samples of each method. The dark bias procedure exhibits the best power stabilization according to Eq. (1) with 75% of the tested samples having values below 2%. The second and third best method was the rapid annealing and illumination with 37.5% and 33% of the tested samples having values below 2%, respectively. The typical number of stabilization cycles for power stabilization was between 3 and 6 cycles (with extreme numbers of 2 to 9) of rapid annealing, dark electric bias and illumination with the power losses as given above.

Based on Figs. 6, 9, 12, and 13 it can be concluded that routes R1, D4, and L3 are the best single routes in the rapid annealing, dark electric bias, and illumination method respectively, considering the low degradation and good stabilization behavior. It should be noted though, that in the rapid annealing method, route R3 has a slightly better stabilization than R1, but much worse degradation rate. In general, the dark bias method would have an advantage over the other two methods due to the absence of illumination, hence all superposing effects of illumination are suppressed (Deceglie et al., 2015). While it worked quite well for the CZTSSe devices of this study, it also shows a high variation of results as stated above in the literature. These arbitrary results show that better standard procedures are still needed to improve the harmonization of the power rating measurements.

## 4. Conclusion

Three different methods to stabilize CZTSSe solar devices have been compared in this study, namely: rapid annealing (R), dark electric bias (D), and illumination (L). Based on the stabilization criteria in the standard IEC 61215-2, routes R3 (100 °C in dark 48 h; followed by 2x illumination 4 h at 25 °C, 1000 Wm$^{-2}$ and $V_{OC}$), D4 (dark bias at 1/3 $I_{SC}$, 50 °C, followed by dark bias at 1/3 $I_{SC}$, 65 °C, 30 min each), and L3 (test requirement of IEC 61215-2 at MPP but at 25 °C) are the best parameter sets with the average $P_{MPP}$ variation of 1.9%, 0.96%, and 1.79% respectively.

The rapid annealing procedure, mainly the R3, may be developed further, e.g., by increasing the annealing temperature above 100 °C. Dark bias stabilization is less time-consuming and can be done without a special, expensive setup, making it an attractive alternative for industry and research. The illumination method could also offer potential for improving measurement procedures by development of route L3 at 25 °C since its power stabilization and performance can compete with the standardized route of L1 at 50 °C.

### Declaration of competing interest

The authors declare that they have no known competing financial interests or personal relationships that could have appeared to influence the work reported in this paper.

### Acknowledgments

This research was funded by the Austrian Institute of Technology (AIT GmbH) and the Austrian Federal Ministry of Education, Science, and Research (BMBWF) via Österreichischer Austauschdienst (OeAD GmbH). WA and AM acknowledge support from the Vienna Doctoral School in Physics.